\documentclass[aps,twocolumn,showpacs,prl,superscriptaddress]{revtex4}

\usepackage{color}
\usepackage{graphicx}

\newcommand{\figurewidth}{\columnwidth}

\def\bec{\beta_{\rm c}}
\def\Tc{T_{\rm c}}

\begin{document}

\title{Extended Scaling Scheme for Critically Divergent Quantities 
in Ferromagnets and Spin Glasses}
\author{I.~A.~Campbell}
\affiliation{Laboratoire des Collo\"ides, Verres et Nanomat\'eriaux, 
Universit\'e Montpellier II, 34095 Montpellier, France}

\author{K. Hukushima}
\affiliation{
Department of Basic Science, University of Tokyo, Tokyo, 153-8902, Japan}

\author{H. Takayama}
\affiliation{
Institute for Solid State Physics, University of Tokyo, 
Kashiwa-no-ha 5-1-5, Kashiwa, 277-8581, Japan}

\date{\today}

\begin{abstract}
 From a consideration of high temperature series expansions in
 ferromagnets and   in spin glasses, we propose an extended scaling
 scaling scheme involving a set of scaling formulae which express to
 leading order the temperature ($T$) and the system size ($L$)
 dependences of thermodynamic observables over a much wider range of $T$
 than the corresponding one in the conventional scaling scheme. The
 extended scaling, illustrated by data on the canonical $2d$ ferromagnet
 and on the $3d$ bimodal Ising spin glass, leads to consistency for the
 estimates of critical parameters obtained from scaling analyses for
 different observables.  
\end{abstract}

\maketitle

Critical divergences of thermodynamical quantities $F(T)$ at continuous
phase transitions are conventionally quoted in terms of the
normalized scaling variable $t = (T-\Tc)/\Tc$, i.e. in the form 
\begin{equation}
F(T)\simeq A_{F}[(T-\Tc)/\Tc]^{-\rho},
\label{eq:standard}
\end{equation}
where $\Tc$, $\rho$ and $A_F$ are the transition temperature, the
critical exponent, and the critical amplitude, 
 respectively. There exist
associated finite size scaling (FSS) rules. It has been clearly
underlined (see e.g.~[\onlinecite{barber:83}]) that this representation
is valid only in the immediate vicinity of $\Tc$, which is a very
restrictive condition both for numerical simulations and experiments. In
particular for the analysis of finite size numerical data on complex
systems such as Ising spin glasses (ISGs), where simulations have many
intrinsic limitations, it is difficult to analyze data while complying
strictly to this condition.

Other scaling variables can be used. In all modern theoretical and
numerical analyses on ferromagnets, e.g., 
Ref.~[\onlinecite{garthenhaus:88,butera:02}], as well as in 
some experimental analyses, e.g.,
Ref.~[\onlinecite{souletie:83,prejean:88}], the scaling variable 
$\tau= 1-\beta/\bec \equiv (T-\Tc)/T$ is used instead of $t$.
Although no general rationalization seems to have been published yet
explaining why one scaling variable should be chosen rather than
another, in this Letter, we propose a coherent scaling scheme for
critically divergent quantities derived from a systematic consideration
of high temperature series expansions (HTSE) which naturally leads us to
use the variable $\tau$. 
Beside this, the HTSE analysis leads us to properly define singular
terms of interest in such a way that they themselves reproduce
appropriate temperature dependence at highest temperatures, i.e., in the
limit $\beta \rightarrow 0$. 
We call this the {\it extended} scaling scheme which we demonstrate
below to be quite powerful at temperatures close to $\Tc$ where 
in practice one makes critical analyses for divergent quantities.

For a ferromagnet and in particular an Ising ferromagnet (IF), our
extended scaling scheme is explicitly described as follows.
\\
\noindent 
(i) We use $\tau$ as the normalized scaling variable, and write 
\begin{equation}
\chi(\beta) \simeq A_{\chi}\tau^{-\gamma},
\label{tau_eqn}
\end{equation} 
for the reduced susceptibility $\chi$ following the standard definition
without a prefactor $\beta$, which is in fact consistent with the idea
of our extended scaling.
\\
(ii) Defining the second moment correlation length $\xi$ through 
$ \mu_2 = \sum_r r^{2} \langle S_0 S_r \rangle = 2d\chi\xi^{2}$, with
$d$ the spatial dimension\cite{butera:02}, we write down $\xi$ as 
\begin{equation}
\xi(\beta) \simeq \beta^{1/2}A_{\xi}\tau^{-\nu}. 
\label{xi_interpolation}
\end{equation}
(iii) Using this form of $\xi(\beta)$, we rewrite the FSS ansatz,
$F(L,\beta) \sim L^{\rho/\nu}{\tilde F}[L/\xi(\beta)]$, as 
\begin{equation}
F(L,\beta) \sim [L/\beta^{1/2}]^{\rho/\nu}{\mathcal F}\left[(L/\beta^{1/2})^{1/\nu}(1-\beta/\bec)\right].
\label{fss_ferro}
\end{equation}

For the $2d$-IF, where the confluent corrections to scaling  are known to
be zero~\cite{garthenhaus:88,salas:99,caselle:02}, we demonstrate that
the above critical expressions with the known exact values of the
critical parameters reproduce $\chi(\beta)$ and $\xi(\beta)$ to a good
approximation right up to high temperatures, and that our FSS form
Eq.~(\ref{fss_ferro}) holds to a high approximation over a very much
wider range of $L$ and $T$ than the conventional one.
For the ISG, $\beta^{2}$ replaces $\beta$ throughout in (i), (ii) and
(iii), giving an entirely novel set of expressions. 
The use of our FSS form appropriately modified for the ISG resolves a
longstanding puzzle in the ISG critical analysis, i.e., published
estimates for the critical exponent $\nu$ through $\chi$ scaling and
through $\xi$ (or the Binder parameter) scaling differ by a large 
factor~\cite{kawashima:96,katzgraber:06}. 
We emphasize here that Eqs.~(\ref{tau_eqn})--(\ref{fss_ferro}) above
are leading order expressions of the divergent quantities. 
This does not mean that we neglect even the confluent corrections to
scaling, but that fits of the numerical data examined below to our
extended scaling scheme only with the leading expressions are quite
satisfactory.
We will discuss separately~\cite{ours} that analyses of published high
precision data on canonical ferromagnets using the present extended
scaling scheme actually give estimates of the confluent correction terms
which improve considerably over those from standard analyses.

In standard spin 1/2 ferromagnets the HTSE for the susceptibility
$\chi(\beta)$ is written as     
\begin{equation}
\chi(\beta)=1 + a_{1}\beta + a_{2}\beta^2 + a_{3}\beta^3 +\cdots
\label{high_T_series}
\end{equation}
where $\beta=J/k_{\rm B}T$ with the coupling constant $J$ and 
$k_{\rm B}$ set to unity~\cite{butera:97,butera:02,butera:02a}. 
The asymptotic form of its factors $a_n$ is eventually dominated by the
closest singularity to the origin (Darboux's First
Theorem~\cite{darboux:78}) which in the simplest case is the physical
singularity, i.e.,  
\begin{eqnarray}
[1-\beta/\bec]^{-\gamma} = 1 + {\gamma}\left(\frac{\beta}{\bec}\right) + \frac{\gamma(\gamma+1)}{2}\left(\frac{\beta}{\bec}\right)^2 +\cdots,
\label{darboux}
\end{eqnarray}
with $\bec$ being the inverse critical temperature. One of the
techniques to relate these two expressions is the ratio method, in which
the recurrence relation  
$
a_{n}/a_{n-1} = (1/\bec)(1+(\gamma-1)/n)
\label{ratio_rule}
$
for large $n$ is used~\cite{guttmann:89}. It is therefore natural to
adopt $\tau =1-\beta/\bec$ as the scaling variable in critical analyses
based on the HTSE theory. 

The HTSE for the second moment $\mu_2(\beta)$ introduced above is of the
form~\cite{butera:02}  
\begin{equation}
\mu_2(\beta) = (b_{1}\beta)[1+ (b_{2}/b_{1})\beta +
 (b_{3}/b_{1})\beta^{2} + \cdots ]. 
\label{mu2-series-mod}
\end{equation}
It diverges at $\Tc$ as $[T-\Tc]^{-(\gamma+2\nu)}$. 
Then, invoking again Darboux's theorem to link the series within the
brackets $[\cdots]$ to the critical divergence, the appropriate extended
scaling form can be written as 
\begin{equation}
\mu_2(\beta)\simeq \beta A_{\mu}\left(1-\beta/\bec\right)^{-(\gamma+2\nu)}.
\label{mu2-scaling}
\end{equation} 
Combined this with Eq.~(\ref{tau_eqn}) for $\chi(\beta)$, 
Eq.~(\ref{xi_interpolation}) for $\xi(\beta)$ is derived. With this
expression for $\xi$ the FSS form becomes Eq.~(\ref{fss_ferro}). 
At the limit $\beta \rightarrow 0$, Eqs.~(\ref{tau_eqn}),
(\ref{xi_interpolation}) and (\ref{mu2-scaling}) have the same $\beta$
dependence as the leading terms of the corresponding HTSE. 
This implies that they merge smoothly to the analytic corrections to
scaling to yield the proper expressions at highest temperatures.
Similar expressions are expected also for confluent corrections to
scaling if they exist.

\begin{figure}[h]
\resizebox{\figurewidth}{!}{\includegraphics{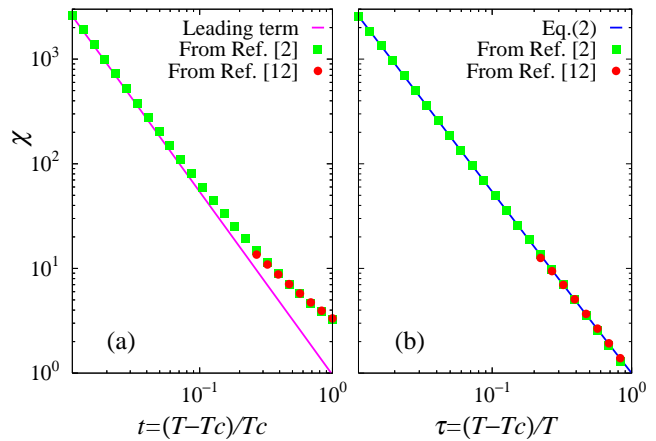}}
\caption{Susceptibility $\chi$ of the $2d$-IF as a function  of log($t$) 
(a) and log($\tau$) 
(b). 
}
\label{fig:2dIFe-chi}
\end{figure}

We exhibit in Figs.~\ref{fig:2dIFe-chi}(a) and \ref{fig:2dIFe-chi}(b)
log-log plots of the susceptibility $\chi$ of the canonical $2d$-IF in the
thermodynamic limit, plotted against log($t$) and log($\tau$),
respectively. 
The data points in the figure are the high-precision results of the
critical~\cite{garthenhaus:88} and HTSE~\cite{butera:02a} analyses. 
The line in Figs.~\ref{fig:2dIFe-chi}(a) and (b) is the power-law expression
of $\chi$ as a function of $t$ and $\tau$, respectively, with $\gamma=7/4$ and the
critical amplitude $A_\chi=0.962581...$\cite{garthenhaus:88}. 
By the $\tau$ scaling, all the data points nearly up to $\tau=1$, i.e.,
to almost $T=\infty$, lie on the scaling expression of
Eq.~(\ref{tau_eqn}) within the accuracy of the figure.  
This result reflects the weakness of corrections to scaling in this
system as mentioned above. 
On the other hand, the deviation of the true $\chi$ data from the $t$
scaling line,  $\chi=A_\chi t^{-\gamma}$, is significant already at,
say, $t\simeq 0.2$ (or $T\simeq 1.2\Tc$).
Thus if the scaling variable $t$ rather than $\tau$ were used for a
critical analysis on this system, there would appear to be very strong
``correction'' terms, which would pollute the evaluation of the $\gamma$
value. 

\begin{figure}[b]
\resizebox{\figurewidth}{!}{\includegraphics{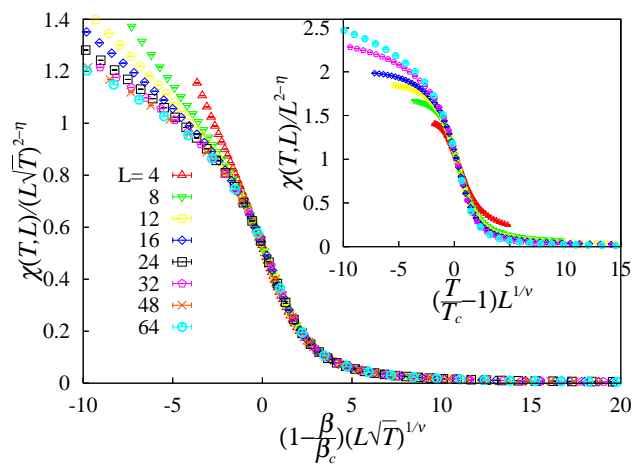}}
\caption{An extended FSS plot for the $2d$-IF susceptibility
 $\chi$.  The points are data obtained by our Monte Carlo (MC)
simulation  at temperatures $0.46 \le T/\Tc \le 1.86$. The
 inset presents a standard FSS plot for the same 
 $\chi$. For both plots the exact scaling parameters are used.}
\label{fig:2dchi-FSS}
\end{figure}

\begin{figure}[t]
\resizebox{\figurewidth}{0.7\figurewidth}{\includegraphics{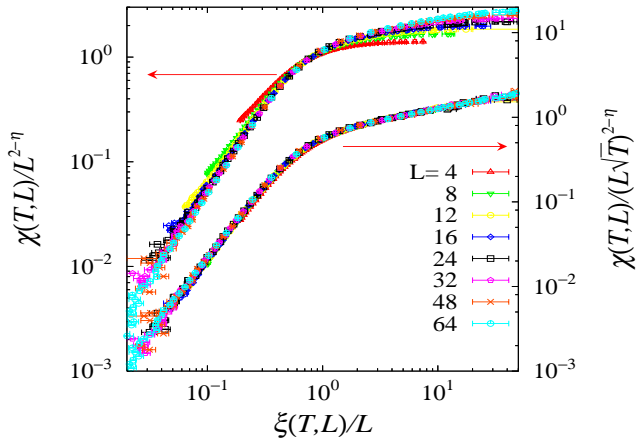}}
\caption{An extended (conventional) FSS plot for the $2d$-IF susceptibility 
$\chi$ normalized by $(L\sqrt{T})^{2-\eta}$ ($L^{2-\eta}$) as a
 function of $\xi(T,L)/L$.}
\label{fig:xikai}
\end{figure}

Figure \ref{fig:2dchi-FSS} shows the FSS plots for the $2d$-IF
susceptibility, the standard one in the inset and the extended one in
the main frame. 
It is clear that the standard form, as expected, gives acceptable
scaling only extremely close to $\Tc$, while the extended form gives
high quality scaling for all temperatures above $\Tc$ examined. 
In Fig.~\ref{fig:xikai} we show the conventional and extended FSS plots
of $\chi(T,L)$ as a function of $\xi(T,L)/L$. 
Note that, since the value $\Tc$ is not involved in this analysis at all
(also in Fig.~\ref{fig:chiSG-xi} below), one can judge straightforwardly
our proposal (iii) by this comparison. 
The consequence is that the extended FSS plot is definitely better than
the conventional one. 
We thus conclude that, at least for the $2d$-IF, our extended scaling
scheme does indeed work much better than the conventional 
one.

\begin{figure}[b]
 \resizebox{\figurewidth}{!}{\includegraphics{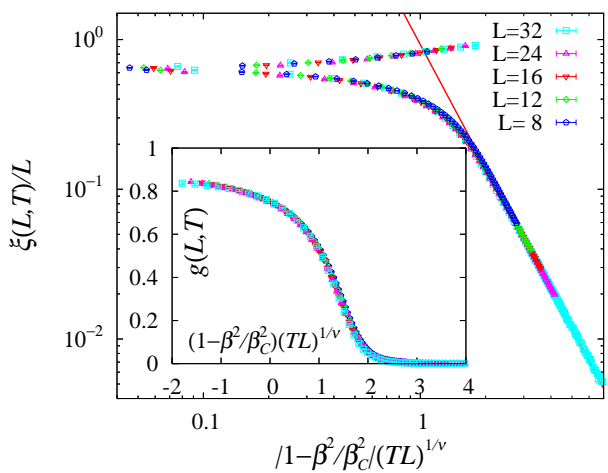}}
\caption{An extended FSS plot of $\xi(L,T)/L$ of $3d$-ISG. The straight
 line with slope 2.72 ($=\nu$) represents an expected asymptotic
 form of the scaling function. The inset  shows the FSS plot of the
 Binder parameter.
}
\label{fig:xi-raw-scal4}
\end{figure}

Let us next discuss an extension of our extended scaling scheme to
the Edwards-Anderson $3d$-ISG model having a symmetric interaction
distribution with zero mean.  
Because of the symmetry, as stated by Daboul {\it et al}~\cite{daboul:04},
only even powers of $\beta$ enter into the HTSE for thermodynamic 
quantities, and the HTSE for the reduced spin glass (SG) susceptibility 
$\chi_{\rm SG}$ (the ordinary one multiplied $T^2$) is of the form
\begin{equation}
\chi_{\rm SG}(\beta) = 1 + c_{1}(\beta^2) + c_{2}(\beta^2)^2 + c_{3}(\beta^2)^3
 +\cdots .
\label{chi_sg}
\end{equation}
Hence once again invoking the Darboux theorem, but this time with
$(\beta/\bec)^2$ replacing $\beta/\bec$, or $\tau'=1-(\beta/\bec)^2$, we
adopt the following scaling form for 
$\chi_{\rm SG}(\beta)$~\cite{daboul:04}
\begin{equation}
\chi_{\rm SG}(\beta) \simeq A_{\chi_{\rm SG}}\left(1 - (\beta/\bec)^2\right)^{-\gamma}.
\label{sg_chi_interpolation}
\end{equation} 
The $\mu_2$ in spin glasses, again due to the symmetry, can be expressed
as even powers of $\beta$ starting from the $\beta^2$ term, though the
coefficients have not been explicitly evaluated yet. 
Hence the scaling form for $\xi(\beta)$ in spin glasses can be taken to
be   
\begin{equation}
\xi(\beta) \simeq \beta A_{\xi_{\rm SG}}\left(1-(\beta/\beta_c)^2\right)^{-\nu}. 
\label{sg_xi_interpolation}
\end{equation}
Then the extended FSS for ISGs can be written as
\begin{equation}
F(L,\beta) \sim [L/\beta]^{\rho/\nu}{\mathcal F}\left[(L/\beta)^{1/\nu}\left(1-(\beta/\bec)^2\right)\right].
\label{fss_sg}
\end{equation}

In contrast to the $2d$-IF, analytical theories are very limited for
ISGs. 
Daboul et al~\cite{daboul:04} were able to make accurate estimates of
$\bec$ and $\gamma$ of the models by the HTSE method but only in
dimension 4 and above.  
We therefore compare our extended $\tau$ scaling scheme with the
conventional one with $t$ variable, without introducing corrections to
scaling in either analysis.
The numerical data used are obtained on the $3d$-ISG system with bimodal
interactions by the exchange MC method.

\begin{figure}[b]
 \resizebox{\figurewidth}{!}{\includegraphics{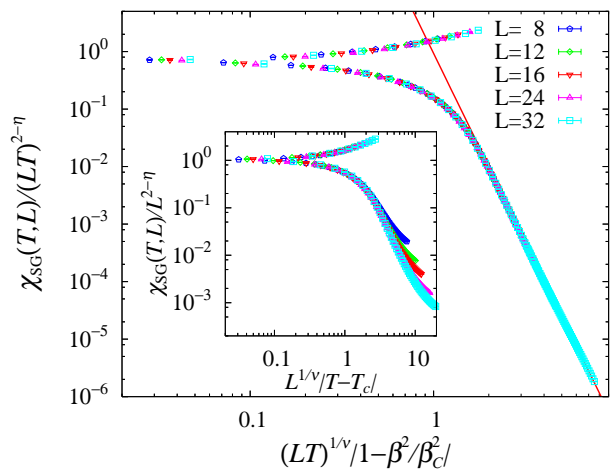}}
\caption{An extended FSS plot of $\chi_{\rm SG}$. The inset is the
 corresponding conventional FSS plot.
}
\label{fig:extended-chiSG-scal}
\end{figure}

In Fig.~\ref{fig:xi-raw-scal4} we show an extended FSS plot for the
correlation length $\xi(L,T)/L$ as a function of the scaling
variable $x=|\tau'|(TL)^{1/\nu}$ based on Eq.~(\ref{fss_sg}). 
In this plot we fix $\Tc=1.11$, which is the optimal value from our
analyses and is consistent with that of Ref.~\cite{katzgraber:06}, and
adjust $\nu$ to obtain the best scaling fit. 
We end up with $\nu=2.72(8)$. 
The fit is surprisingly good for all the data with $L$'s indicated in
the figure and at $T$ from $0.81T_c$ to $7.2T_c$. 
The slope of the straight line in the range $|x| \gg 1$ is 2.72
($=\nu$).   
Note that data points on the line are not only those at
sufficiently high temperatures with $\xi \simeq \beta$ from
Eq.~(\ref{sg_xi_interpolation}) but also those in the
critical range with $\beta \ll \xi \ll L$. 

Figure~\ref{fig:extended-chiSG-scal} shows an extended FSS plot for 
$\chi_{\rm SG}$. This plot is obtained by fixing $\Tc=1.11$ and
$\nu=2.72$ and by adjusting $\eta$ to give $\eta=-0.40(4)$. 
We obtain quite satisfactory scaling for all our data. 
In contrast, a conventional FSS plot using the variable $t$ as shown in
the inset is rather poor except for the immediate vicinity of $x=0$. 
The fit yields a small apparent value of $\nu = 1.47(3)$ similarly to
the previous conventional FSS analyses~\cite{kawashima:96,katzgraber:06}. 
In Fig.~\ref{fig:chiSG-xi} we demonstrate conventional and extended
scaling plots for $\chi_{\rm SG}$ versus $\xi$. 
The comparison of the two implies that our scaling scheme with (ii) and
(iii) is definitely more appropriate also for the $3d$-ISG. 

\begin{figure}[t]
 \resizebox{\figurewidth}{0.65\figurewidth}{\includegraphics{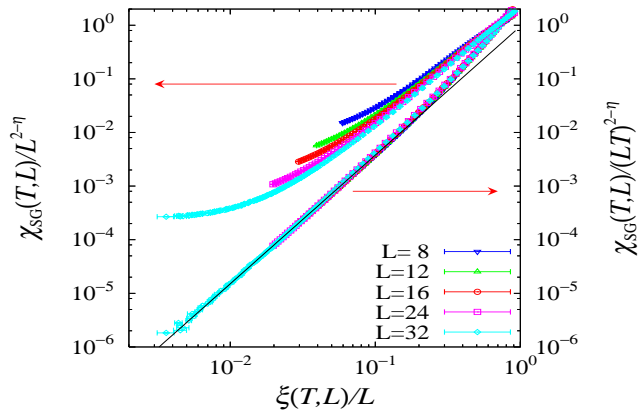}}
\caption{An extended (conventional) FSS plot for the $3d$-ISG susceptibility 
$\chi_{\rm SG}$ normalized by $(LT)^{2-\eta}$ ($L^{2-\eta}$) as a
 function of $\xi(T,L)/L$.}
\label{fig:chiSG-xi}
\end{figure}

One can remark that our extended scaling scheme with only the leading
term for each divergent observable gives high quality fit over the
entire range of $L$ and $T$ examined. 
Another important result here for the SG study is that the extended FSS
analyses on $\xi$, $g$ and $\chi_{\rm SG}$ with the implicit assumption
that corrections to scaling are weak yield a unique critical parameter
set. 
This is in sharp contrast to standard FSS methods for which the estimate
for $\nu$ obtained from $\chi$ scaling with the same assumption is
considerably smaller than that from $\xi$ or $g$ 
scaling~\cite{kawashima:96,katzgraber:06}. 
In this context, we note that numerical data on the same $3d$-ISG model
have also been analysed using $t$ scaling together with strong
correction to scaling terms~\cite{palassini:99,ballesteros:00}.
We have checked that our data can equally well be analysed using a very
similar method to that of Ref.~[\onlinecite{palassini:99}], and the
excellent fit to the data from the extended scaling without corrections
to scaling is obtained with two fewer fitting parameters. 
This does not imply that corrections to scaling are absent, but just
that their influence on the fits examined here is rather weak.

In conclusion, by considering the intrinsic structural form of high
temperature series developments, we have proposed an extended scaling
scheme with appropriate scaling expressions for thermodynamic
observables in ferromagnets and in spin glasses, and have demonstrated
the results which support it strongly. 
One of them is the direct comparison of our extended scaling on $\chi$
vs. $\xi$ with that of the conventional one. 
Another is the result that, within our scheme, the leading order
critical power-law expressions with a coherent set of critical
parameters remain good approximations to the true behavior over 
a much wider temperature range than with the standard $t$ scaling.  
From these results we consider that our extended scaling scheme with the
variable $\tau$ is more fundamental than the conventional $t$ scaling. 

We would like to thank P.~Butera for all his painstaking and
patient advice, H.~Katzgraber for numerous and helpful remarks,
and C.~Chatelain for pointing out that the squares in the ISG series
arise from the interaction symmetry. 
This work was supported by the Grants-In-Aid for Scientific Research
(No.~14084204 and No.~16540341) and NAREGI Nanoscience project, both
from MEXT of Japan. 
The numerical calculations were mainly performed on  the SGI Origin
2800/384 at the Supercomputer Center, ISSP, the University at Tokyo.

\end{document}